\documentclass[11pt]{article}
\usepackage{latexsym} \usepackage{epsf}
\usepackage{a4}
\usepackage{amsfonts}
\usepackage{amsmath}
\usepackage{amsthm}
\usepackage{amssymb}
\renewcommand{\theequation}{\thesection.\arabic{equation}}
\newcounter{subequation}[equation]
\makeatletter

\expandafter\let\expandafter\reset@font\csname reset@font\endcsname

\def\subeqnarray{\arraycolsep1pt
    \def\@eqnnum\stepcounter##1{\stepcounter{subequation}%
        {\reset@font\rm(\theequation\alph{subequation})}}
\jot5mm     \eqnarray}

\makeatother

\def\be{\begin{equation}}
\def\ee{\end{equation}}
\def\bea{\begin{eqnarray}}
\def\eea{\end{eqnarray}}
\def\ba{\begin{array}}
\def\ea{\end{array}}
\def\dd{\partial}

\def\half{\frac{1}{2}}

\def\one#1{#1^{\raise5pt\hbox{$\scriptstyle\!\!\!\!1$}}\,{}}
\def\two#1{#1^{\raise5pt\hbox{$\scriptstyle\!\!\!\!2$}}\,{}}

\def\tilde{\widetilde}
\def\II{\hbox{{1}\kern-.25em\hbox{l}}}
\makeatletter
\def\binrel@#1{\begingroup
  \setboxz@h{\thinmuskip0mu
    \medmuskip\m@ne mu\thickmuskip\@ne mu
    \setbox\tw@\hbox{$#1\m@th$}\kern-\wd\tw@
    ${}#1{}\m@th$}%
  \edef\@tempa{\endgroup\let\noexpand\binrel@@
    \ifdim\wdz@<\z@ \mathbin
    \else\ifdim\wdz@>\z@ \mathrel
    \else \relax\fi\fi}%
  \@tempa
}
\let\binrel@@\relax
\def\overset#1#2{\binrel@{#2}%
  \binrel@@{\mathop{\kern\z@#2}\limits^{#1}}}
\def\underset#1#2{\binrel@{#2}%
  \binrel@@{\mathop{\kern\z@#2}\limits_{#1}}}
\makeatother
\newfont{\bbd}{msbm10 scaled\magstep1}

\newtheorem{prop}{Proposition}

\begin{document}

\begin{center}
{\LARGE Iterative construction of $U_q (s\ell (n+1)) $ \\
representations and Lax matrix factorisation } 

\vspace{1cm}

{\large  S. Derkachov$^{a}$
\footnote{e-mail:derkach@euclid.pdmi.ras.ru},
D. Karakhanyan$^b$\footnote{e-mail: karakhan@lx2.yerphi.am},
R. Kirschner$^c$\footnote{e-mail:Roland.Kirschner@itp.uni-leipzig.de} \\
[3mm]  and P. Valinevich$^d$ 
}

\begin{itemize}
\item[$^a$]
St. Petersburg Department of Steklov Mathematical Institute
of Russian Academy of Sciences,
Fontanka 27, 191023 St. Petersburg, Russia.
\item[$^b$]
Yerevan Physics Institute, \\
Br. Alikhanian st. 2, 375036 Yerevan, Armenia.
\item[$^c$]
 Institut f\"ur Theoretische
Physik, Universit\"at Leipzig, \\
PF 100 920, D-04009 Leipzig, Germany
\item[$^d$] 
St. Petersburg State University, 198504 St. Petersburg, Russia. 
\end{itemize}

\vspace{1cm} \noindent {\bf Abstract}
\end{center}

\noindent
The construction of a generic representation of 
$g\ell(n+1)$
or of the trigonomentric deformation of its enveloping 
algebra known as algebraic induction is conveniently 
formulated in terms
of Lax matrices. The Lax matrix of the constructed 
representation factorises into parts determined by the 
Lax matrix of 
a generic representation of the algebra with reduced rank
and others appearing in the factorised expression of the
Lax matrix of the special Jordan-Schwinger representation.

%%%%%%%%%%%%%%%%%%%%%%%%%%%%%%%%%%%%%%%%%%%%%%%%%%%%%%%%%%%%%%%%%%%%%%%%%%%%%%
%{\small \tableofcontents}
\renewcommand{\refname}{References.}
\renewcommand{\thefootnote}{\arabic{footnote}}
\setcounter{footnote}{0}

\vspace{1cm}

\section{Introduction}

Representations of $g\ell (n+1)$ as well as of the trigonometric deformation
$U_q (g\ell (n+1)) $ of its enveloping algebra can be obtained
from representations of the corresponding algebras with rank smaller by one
unit and a set of $n+1$ Heisenberg  pairs $x_i, \dd_i, i= 1, ..., n+1$.
This iterative procedure is known as algebraic induction method.
 The background of this method is the general method of
induced representations, in particular the construction of $U(n)$
representations from their characters \cite{GN, BW}.
Biedenharn and Lohe \cite{BL} developed the method of 
algebraic induction
in application to quantum groups, relying on earlier results
\cite{H}
and going back to Holstein and Primakoff \cite{HP}.
 Representations constructed by this method in terms of 
Heisenberg operators have been considered in 
\cite{Dobrev:1994ne}.

The $RLL$ relation with the Lax matrix $L$ composed of the
generators of the considered algebra is a simple and 
compact expression of the algebra relations.
We shall show that the Lax matrices provide a natural 
formulation of the
algebraic induction, allowing an easy derivation and simplifying essentially
the expression of the constructed $g\ell (n+1)$ repesentation 
in terms of a $g\ell (n)$ representation
and a set of $n+1$ Heisenberg pairs $x_i, \dd_i$.
The latter are used to construct first of all 
the special Jordan-Schwinger form
of representations of $g\ell (n+1)$ and $U_q (g\ell (n+1))$.
The corresponding Lax matrices have simple factorization 
properties and have a point of degeneracy at the spectral 
parameter value $u=1$. The matrix product of Lax matrices
of two representations as well as the Lax matrix of the 
tensor product representation obey the Yang-Baxter $RLL$
relation. Then choosing one of the representations to be of the
Jordan-Schwinger form we observe that constraints can be imposed
reducing the other tensor factor to a   
$g\ell (n+1)$ or $U_q (g\ell (n+1))$ representation without
disturbing the algebra relation for the tensor product 
generators. Underlying relations appear similar to the
classical fusion method \cite{Kulish:1981gi}.
This results in the algebraic induction,
i.e. the construction of representations of 
$g\ell (n+1)$ and $U_q (g\ell (n+1))$
in terms of a representation of 
$g\ell (n)$ or $U_q (g\ell (n))$ and $n+1$ 
Heisenberg conjugated pairs. Further, the relation between the
product of Lax matrices of the two representations and the
tensor product Lax matrix results in triangular factorisation
relations for the latter Lax matrix. This factorised expression
provides a compact formulation of the algebraic induction.

Our motivation of reconsidering the algebraic induction method in
relation to Lax matrices arises from the task of factorisation
of the Yang-Baxter R-operator acting on the tensor product
of two generic representations.
In the case of  of $s\ell (2)$ this factorisation has been
established in
\cite{Derkachov:2005hw}
by regarding the action of $R$ in the $RLL$ relation as the
permutation of pairs of parameters built from the representation parameter
$\ell$ and the spectral parameter $u$ by decomposing 
this permutation into more elementary ones.
Solving the defining conditions for the factor operators
is essentially simpler compared to the conditions for the
complete R-operator.
The method has been developed in application to the
trigonomentric and elliptic deformations of $s\ell (2)$
\cite{Derkachov:2007gr},
to $s\ell (3)$ and its trigonometric deformation 
and also to $s\ell (n)$ \cite{Derkachov:2006fw}.
 All cases rely on triangular factorization
relations for the corresponding Lax matrices.

The triangular factorization of the $s\ell (n)$ 
Lax matrix has been obtained
in \cite{Derkachov:2006fw} by 
using the representation induced from the Borel subgroup of triangular
matrices. An extension this approach to factorization 
to the trigonometric deformation case may be allowed using a
formulation like in \cite{Awata:1993hj,Dobrev:1993sr}.

We start from the tensor product of two repesentations of 
$g\ell (n+1)$ or $U_q (g\ell (n+1))$ and
consider the relation of the Lax matrix composed of the co-product
generators to the product of Lax matrices of the tensor factors
(Sect. 2). Then we analyse the factorisation properties of the
Lax matrix of the special Jordan-Schwinger representations, where the
$g\ell (n+1)$ algebra generators are constructed from
the mentioned Heisenberg conjugated pairs (Sect. 3).  The first of the tensor
factors
is substituted in the Jordan-Schwinger form. The second
tensor factor can
be constrained from $g\ell (n+1)$ to $g\ell (n)$ without disturbing the
algebra relations in the tensor product (Sect. 4).
The constrained tensor product Lax matrix is expressed as a 
product of factors of the
Jordan-Schwinger Lax matrix and a  matrix involving essentailly
the $g\ell (n)$ generators (Sect. 5). 
Together with the factorization
properties of the Jordan-Schwinger Lax matrix this leads to
the wanted factorisation formulae and a compact formulation
of the algebraic induction.

\section{Tensor product in terms of Lax matrices}
\setcounter{equation}{0}

We consider  the co-product of the $U_q (g\ell(n+1))) $ algebra.
One of the factors will be later substituted
in the restricted Jordan-Schwinger form.
We define the co-product on the generators in the
Chevalley form, $e_i, f_i, h_i =
\half (N_i - N_{i+1}), i=1, ..., n$ in the symmetric way,

\be 
 \Delta (e_i) =  e_i\otimes q^{\half (N_{i+1} - N_i ) }
+ q^{\half (N_{i} - N_{i+1} ) }\otimes e_i, \ \ \ee
$$
\Delta (f_i) = f_i\otimes q^{\half (N_{i+1} - N_{i}) } +
q^{\half
(N_{i} - N_{i+1})}\otimes f_i, \ \ \Delta (N_i) =
N_i\otimes\II +
\II\otimes N_i.
$$
Having in mind the representations $\pi^{(1)}, \pi^{(2)}$ on
linear spaces $V^{(1)}, V^{(2)}$ and also the tensor product
$V^{(1)} \otimes V^{(2)}$ we shall use the 
notation by subscripts $(1), (2), (12)$
(called Sweedler s notation in \cite{KS}) and omit
the symbol $\otimes$.  
\be
\label{co}
 \Delta (e_i) = e_i^{(12)} =  e_i^{(1)} q^{\half (N_{i+1}^{(2)} - N_i^{(2)} ) }
+ e_i^{(2)} q^{\half (N_{i}^{(1)} - N_{i+1}^{(1)} ) }, \ \
\ee
$$
\Delta (f_i) = f_i^{(12)} 
= f_i^{(1)} q^{\half (N_{i+1}^{(2)} - N_{i}^{(2)}) }
+ f_i^{(2)} q^{\half (N_{i}^{(1)} - N_{i+1}^{(1)}) },
$$ $$\Delta (N_i) = N_i^{(1)} + N_i^{(2)}.
$$
The Cartan-Weyl generators are defined iteratively as
\bea
\label{iter}
E_{i,i+1} = e_i, \ \ \ E_{i+1,i} = f_i, \qquad\qquad\qquad\cr
E_{ij} = [E_{i, j-1}, E_{j-1,j} ]_{q}, \qquad i+1<j, \cr 
E_{ij} =
[E_{i, i-1}, E_{i-1,j} ]_{q^{-1}}, \qquad i>j+1. 
\eea
Here we use an appropriate modification of the
commutator notation
defined  as $[A, B]_{q} = A B - q B A $.

We intend to write the co-product explicitly in the latter basis.
The result can be compactly formulated in terms of the upper and lower
triangular parts of the Lax matrices.
Let us  compose the Lax matrix according to
Jimbo \cite{Jimbo:1985zk},
\be
\label{Lax}
L_{ij} (u) = q^{-(u-\half) - \half (E_{ii} + E_{jj})}
E_{j,i} , \qquad i> j \ee
$$ L_{ij} (u) = q^{+(u-\half) + \half (E_{ii} + E_{jj})}
E_{j,i}, \qquad i< j $$
$$ L_{ii} (u) = [ u + E_{ii} ] $$
and consider the standard decomposition 
\be 
\label{Laxpm} 
\lambda \ L(u) = q^u L_+ - q^{-u} L_- \ee 
where $\lambda = q - q^{-1} $ and
$(L_+)_{ij} =0$, $i>j$, $(L_-)_{ij} =0$, $i<j$. We use the standard
notation $[x] = (q^x - q^{-x}) \lambda^{-1} $.

Consider now the generators and Lax matrices for two representations
of the algebra, $E^{(1)}_{ij}, E^{(2)}_{ij},$ $
L^{(1)} (u), L^{(2)} (u) $.
Then the Lax matrix composed by the same rule (\ref{Lax})
with the
co-product Cartan-Weyl generators has the form
$$ \lambda L^{(12)}(u) = q^u L^{(12)}_+ - q^{-u} L^{(12)}_- $$
where
\be
\label{Lax12}
L^{(12)}_{\pm} = L^{(1)}_{\pm} \ L^{(2)}_{\pm}
\ee
A proof of this known relation  
\cite{Faddeev:1987ih,Faddeev:1996iy,Isaev:1995cs} 
is given in Appendix A. 

Let us recall also the situation in the undeformed case.
$$  L_{ij}(u) = u \delta _{ij}  +  E_{ji} $$
Here the Lax matrix for the tensor product is composed according to the 
latter
prescription with the trivial co-product
$$ \Delta_1 (E_{ij}) = E^{(12)}_{ij} =
E_{ij}^{(1)} + E_{ij}^{(2)}, $$
$$ L^{(12)}_{ij} (u) = u \delta_{ij} +
E^{(1)}_{ji} + E^{(2)}_{ji}. $$

Consider now the Yang-Baxter relation involving the fundamental
$(n+1)\times(n+1)$ R-matrix and the Lax matrices \be \label{RLL}
 \check R(u-v) L_1(u) L_2 (v) = L_1 (v) L_2 (u) \check R(u-v),
\ee
$$ L_1 = L \otimes I , \ \ L_2 = I \otimes L. $$
This relation is also fulfilled if one substitutes the Lax matrix
by the matrix product
$L^{(1)} (u+ \delta_1) \ L^{(2)} (u+ \delta_2)$
or by the Lax matrix composed from the co-product generators
$ L^{(12)} (u) $, eqs. (\ref{co}, \ref{Lax12}).
\begin{prop}
In both the rational (undeformed)  and the trigonometric cases the
following relation holds for the Lax matrices of representations
$\pi^{(1)}, \pi^{(2)}$ and of the tensor product 
$\pi^{(12)}$
$$ L^{(1)} (u+ \delta_1) L^{(2)} (u+ \delta_2) = [u+ \delta_3] \
L^{(12)} (u+ \delta_4) +  L^{(1)} ( \Delta_1) L^{(2)} ( \Delta_2) $$
with the shifts related as
\be
\label{relp}
 \delta_1 + \delta_2 = \delta_3 + \delta_4,\qquad
\Delta_1 + \Delta_2 = - \delta_3 + \delta_4,\qquad \Delta_1 -
\Delta_2 = \delta_1 - \delta_2 \ee
\end{prop}
The proof is  straightforward in both the undeformed and deformed cases by
substituting the explicit forms of the Lax matrices and comparing terms with
the same dependence on the spectral parameter $u$.

\section{ Jordan-Schwinger representations}
\setcounter{equation}{0}

 We construct generators of
$g\ell (n+1)$ by taking $n+1$ Heisenberg pairs $x_i, \dd_i,
i= 1, ..., n+1$. In the undeformed case

$$ E_{ij} = x_i \dd_j $$
obey the Lie algebra relations. The constraint
\be
\label{repc}
\sum x_i \dd_i = 2 \ell
\ee
can be imposed to fix a representation of
$s\ell (n+1)$, irreducible for generic $\ell$.
We postpone the elimination of a degree of freedom
by the constraint (\ref{repc}) and discuss its effect
in Appendix B.

In the deformed case we start with
$$ E^{J}_{ij} = {x_i \over x_j} [N_j], \qquad \ N_j = x_j \dd_j
\qquad \ i,j = 1, ..., n+1 $$ We check easily that for $i, j, k$
pairwise different \be \label{EJ}
  [E_{ij}^{J},E_{j k}^{J} ]_{q^{\pm 1}} = q^{\mp N_j}
E_{i k}^{J}, \qquad \
 [E_{ij}^{J},E_{j i}^{J} ]_{1} = [N_i - N_j ] .
\ee

These operators can be related to the Chevalley basis
of  $U_q (sl(n+1) ) $ algebra as
$$
e_i = E_{i,i+1}^J, \qquad f_i = E_{i+1,i}^J, \qquad 2 h_i = [N_i -
N_{i+1}], \qquad i= 1, ..., n
$$
The algebra relations including Serre's relations can be checked.

Alternatively one can extend the construction to the
Cartan-Weyl generators by defining them by q-commutators
iteratively (\ref{iter}).
In our case this leads to
\be
\label{Ex}
 E_{ij} = q^{-(N_{i+1} + ...+ N_{j-1})} E_{ij}^J,\qquad i<j,
\ee
$$ E_{ij} = q^{(N_{i-1} + ...+ N_{j+1})} E_{ij}^J,\qquad i>j.
\qquad \qquad i,j = 1, ..., n+1
$$
The algebra relations in the
Cartan-Weyl form are fulfilled. This property is preserved after
imposing the constraint $ \sum x_i \dd_i = 2 \ell $.

Let us now compose the Lax matrix according to (\ref{Lax})
 substituting the generators in Jordan-Schwinger form,
\be
\label{LaxN}
 L_{ij} (u) = (q^{\pm 1})^{(u-\half) + N_{ij}}
 \ {x_j \over x_i} [N_i],
i \not = j
\ee
$$ L_{ii} = [u+ N_i], $$
The sign $+$ or $-$ in the exponent of $q$ stands in the cases $i<j
$ or $i>j$, respectively. We have introduced the notation $N_{ij}=
N_{ji},\;\; i \not = j$ where for  the case $i<j$
$$ N_{ij}= \half N_i + N_{i+1} + ...+ N_{j-1} + \half N_j  $$

The Lax matrix in the form before imposing the constraint
(\ref{repc})
can be simplified by the following similarity transformation,
\be
\label{DLD}
 D^{(x) }  L (u) D^{(x) -1} = \tilde L (u)
\ee where $D^{(x)}$ is a diagonal matrix with
 \be \label{Dx}
D^{(x )}_{ii} = q^{-N_{i,n+1}} \ x_i, \qquad i\le n,\qquad
D^{(x)}_{n+1,n+1} = x_{n+1}.
 \ee
The simplified Lax matrix $\tilde L$ has the elements
$$ \tilde L_{ij} = (q^{\pm 1})^{u-1} [N_i],
  \qquad \
   \tilde L_{ii} = [u-1 + N_i],  $$
Again the sign $+$ stands in the case $i<j$ and the sign $-$
in the case $i>j$.
\bea
\label{Ltilde}
\tilde L (u)
= \left(
\begin{array}{ccccc}
[u-1 + N_1]  & q^{u-1} [N_1] & ...&
q^{u-1} [N_1] &  q^{u-1}[N_1]  \\
q^{-u+1} [N_2] & [u-1+ N_2] &  ...&
q^{u-1} [N_2] &  q^{u-1} [N_2]  \\
...&...&...&...&... \\
q^{-u+1} [N_{n+1}] &q^{-u+1} [N_{n+1}] & ...&
q^{-u+1} [N_{n+1}] & [u-1+ N_{n+1} ] \\
\end{array} \right )
\eea
Note that u=1 is a singular point
of this matrix.
\bea
\tilde L (1)
= \left(
\begin{array}{ccccc}
[N_1]  &  [N_1] & ...&
 [N_1] &  [N_1]  \cr
[N_2] & [N_2] &  ...&
 [N_2] &   [N_2]  \cr
...&...&...&...&... \cr
[N_{n+1}] & [N_{n+1}] & ...&
 [N_{n+1}] & [ N_{n+1} ] 
\end{array} \right )
= D^N \ M_1, 
\nonumber
\eea

\be
\label{degen}
  L (1) = D^{(x) -1} D^{N} M_1 D^{(x)}.
\ee $M_1$ denotes the $(n+1)\times(n+1) $ matrix with all elements
equal to 1 and $D^N$ denotes the diagonal matrix with $D^N_{ii} =
[N_i]$.

It is not difficult to see that the matrix $\tilde L (u) $ can be
transformed to  an upper triangular matrix  $\tilde K$,
$$ \tilde L = \tilde M_L \ \tilde K \tilde M_R $$
with  special lower triangular matrices $M_L, M_R$ having $1$ on the
diagonal and the only further non-zero elements
$$ (\tilde M_L)_{n+1, i} = - q^{\alpha_i},  \ \ \
 (\tilde M_R)_{n+1, i} =  q^{2(1-u)} $$

Later we shall find it useful to express these matrices in terms of
the standard matrix $m_1 $, 
$$
m_1 =  \left(
\begin{array}{ccccc}
1  &  0 & ...&
 0 &  0  \\
 0 & 1 &  ...&
 0 &   0  \\
...&...&...&...&... \\
 1 &1 & ...&
 1 & 1 \\
\end{array} \right )
$$
with diagonal and last-row elements
equal to 1 and other elements zero and the diagonal matrices
\be
\label{DLR}
( D_L )_{ii}(u)  = q^{\alpha_i}, \ \
( D_R )_{ii}(u)  = q^{2(1-u)}, i \le n,
\ee
$$ ( D_L )_{n+1,n+1}(u) = 1, \ \
( D_R )_{n+1,n+1}(u) = 1.
$$
as
$$
\tilde M_L = D_L^{-1} m_1^{-1} D_L, \ \ \
\tilde M_R = D_R^{-1} m_1  D_R.$$

Let us first write down the matrix $
\tilde L \ \tilde M_R^{-1}  $
$$
\!\!\!\!\!\!\! =\!\! \left(\!\!\!
\begin{array}{ccccc}
[u-1] q^{N_1} & \lambda [u-1] [N_1] &\!\! ...&
\lambda [u-1] [N_1] &  q^{u-1} [N_1]  \\
0 & [u-1] q^{N_2} & \!\! ...&
\lambda [u-1] [N_2] &  q^{u-1} [N_2]  \\
...&...&\!\!...&...&... \\
-[u-1] q^{2(1-u)-N_{n+1}} &\!\!-[u-1] q^{2(1-u)-N_{n+1}} &\!\! ...&
\!\!-[u-1] q^{2(1-u)-N_{n+1}} &\!\! [u-1+ N_{n+1} ] \\
\end{array} \!\!\!\right )
$$
This matrix has vanishing elements below the diagonal
besides of the last row.

$\tilde K $ is obtained by multiplying this matrix
by $ \tilde M_L^{-1}$ from the left in order to clean up the last row. 
Choosing
$$ \alpha_i = 2(1-u) -N_{n+1} - 2 N_1 - ...-2 N_{i-1} - N_i, \qquad
i = 1, ..., n $$
we obtain the wanted form
\be
\label{tildeK}
\tilde K
=
\ee
 $$
\!\!\!\left(\!\!\!
\begin{array}{ccccc}
[u\!-\!1] q^{N_1} &\!\! \lambda [u\!-\!1] \ [N_1] & ...&
\!\!\lambda [u\!-\!1][N_1 ] &\!\! q^{u-1} [N_1]  \\
0 &\!\! [u\!-\!1] q^{N_2} &  ...&
\!\!\lambda [u\!-\!1] [N_2 ] &\!\! q^{u-1} [N_2]  \\
...&...&...&...&... \\
0 &0 & ...& 0 &\!\! [u\!-\!1\!+\!  N_1\! +\! ...\! +\!
N_{n+1} ] q^{-N_1 - ... -N_n} \\
\end{array}\!\!\! \right )
$$
We have obtained the factorized form of the
Jordan-Schwinger Lax matrix
\be
\label{LaxJS}
L(u) = M_L(u) K(u) M_R (u)
\ee
where $M_L, M_R$ are lower triangular with 1 on the
diagonal and the only further non-vanishing elements on the last
row.
$$K (u) = D^{(x) -1} \tilde K(u) D^{(x)} $$
is upper triangular and $\tilde K$ involves on the operators
$N_i,\;\; i=1, ..., n+1$. The left and right lower triangular
factors are calculated from the standard matrix $m_1$ by similarity
transformation with diagonal matrices,
 \be \label{MLR}
M_L (u) = D_L^{-1} (u) m_x^{-1} D_L(u), \ \ M_R (u) = D_R^{-1} (u)
m_x D_R (u), \ \ m_x = D^{(x) -1} m_1 D^{(x)}.
 \ee

Notice that there is an alternative factorized form where lower
triangular matrices appear instead of upper triangular and vice
versa. In the above factorization we have given a distinguished role
to the last column and last row in (\ref{Ltilde}). The mentioned
alternative form is obtained by distinguishing instead the first
column and first row. More forms can be obtained by distinguishing
in the analogous way the row and column of number $i$. Further, one
can start the first step of producing zero elements in
(\ref{Ltilde}) with a row instead of a column. Then the roles of
$M_L$ and $M_R$ are in interchanged. Whereas in the considered form
$M_R$ has simpler elements than $M_L$ this will then appear
oppositely.

%%%%%%%%%%%%%%%%%%%%%%%%%%%%%%%%%%%%%%%%%%%%%%%%%%%%%%%%%%%%
The Jordan-Schwinger form of $g\ell (n+1)$ does not
cover all representations. For some particular values of
$\ell$ the finite-dimensional representations symmetric in the
tensor indices are involved, whereas arbitrary
Young tableaux of index permutation symmetry are not covered
by this form. As explained in Appendix B the representation
modules of lowest weight spanned by polynomials 
have weights of the restricted form with $n-1$ zero components
and one component equal to $\ell$.  
The representation constraint (\ref{repc}) commutes with
 the Lie algebra but not with all generators of the
Heisenberg algebra from which the latter is composed.
We would like to use this constraint to eliminate
$N_{n+1}$ and $x_{n+1}$. Some changes are expected
because $x_{n+1}$ does not commute with the constraint.
In Appendix B we show that this leads to minor modifications of
the resulting factorization. A remarkable point is that
the dependence on the representation parameter $\ell$
introduced by this constraint can be localized in the left
factor $M_L$.

In the undeformed case the corresponding formulae  for the Lax
matrices and the factorization are obtained by taking the limit $q
\rightarrow 1 $. The factorization (\ref{LaxJS}) holds where the factors
on r.h.s. simplify as
$$ M_L \to m_x^{-1} , \ \ M_R \to m_x $$
where in this limit
$$  
m_x \to
\left(
\begin{array}{ccccc}
1  &  0 & ...&
 0 &  0  \\
 0 & 1 &  ...&
 0 &   0  \\
...&...&...&...&... \\
 \frac{x_1}{x_{n+1}} &\frac{x_2}{x_{n+1}} & ...&
 \frac{x_n}{x_{n+1}} & 1 \\
\end{array} \right )
$$
The central factor simplifies to
\bea
\tilde K
\rightarrow
\!\!\!\left(\!\!\!
\begin{array}{ccccc}
u\!-\!1  &\!\! 0 & ...&
\!\!0 &\!\! N_1  \\
0 &\!\! u\!-\!1  &  ...&
\!\!0 &\!\!  N_2  \\
...&...&...&...&... \\
0 &0 & ...& 0 &\!\! u\!-\!1\!+\!  N_1\! +\! ...\! +\!
N_{n+1}   \\
\end{array}\!\!\! \right ).
\eea

\section{Reduction}
\setcounter{equation}{0}

Now we turn to the special case where one of the tensor factors is
constructed by Jordan-Schwinger generators. We shall denote the
corresponding operators by subscript $x$ instead of $(1)$,
$$ E_{ij}^{(1)} = E_{ij}^{x}, \ \ L^{(1)} (u) = L^{x} (u). $$
The second tensor factor is so far
a generic $g\ell (n+1) $ or $U_q (g\ell (n+1) )$ representation.
We shall omit later the label $(2)$.
To prevent confusions we denote in the following
the diagonal generators
of the second factor $E_{ii}= N_i^{(2)}$ by $E_i$
and the diagonal generators of the first $E^{x}_{ii} = N_i^{(1)}$
by $N_i$.

We recall that $u=1$ is a point of degeneracy of $L^{x}(u)$ and
consider the particular case of the relation (\ref{relp})
\be
\label{rel}
 L^{x} (u+ 1) L^{(2)} (u) = [u] \
L^{(12)} (u+ 1) +  L^{x} ( 1) L^{(2)} ( 0) \ee We shall investigate
the condition for vanishing of the remainder $ L_r = L^{x} ( 1)
L^{(2)} ( 0) $, and we shall see that this results in the first step
of the reduction of the second tensor factor from 
$U_q (g \ell (n+1) ) $ to $U_q (g\ell (n) )$ 
preserving the algebra relations of
$U_q (g \ell (n+1) ) $ for the reduced generators involved 
in  the Lax matrix $ L^{(12)} (u) $ for the tensor product
and the Yang-Baxter $RLL$ relations for both 
$L^{x} (u+ 1) L^{(2)}(u) $ and $ L^{(12)} (u) $.
This is reminicent to the fusion procedure 
\cite{Kulish:1981gi},
where in a similar way higher represetations are constructed 
starting from a tensor product and applying a projection
preserving the algebra relations.

\subsection{Undeformed case}

We substitute $L^{x} (1)$ using (\ref{Ltilde})
and establish as a sufficient condition for the vanishing
of the second term on r.h.s. of (\ref{rel})
the vanishing of the following set of $n+1$  operators
\be
\label{phi1}
\varphi_i = \sum_1^{n+1}  x_s E_{i s}
\ee

We would like to see whether the constraints $\varphi_i = 0$ are
compatible with the original algebra relations. This can be done by
analyzing the consequences of the RLL relation for $L$ substituted
as $L^{x} (u+ 1) L^{(2)} (u) $ or as $L^{(12)} (u+ 1) $ and the
relation (\ref{rel}). In the undeformed case it is easier to study
the commutation relations.

Indeed, we have
$$
[ \varphi_i, \varphi_j ] = x_i \varphi_j - x_j \varphi_i
$$
and
\be
\label{ideal}
[ E_{ij}^{x} + E_{ij}, \varphi_k ] = \delta_{jk} \varphi_i
\ee
This shows that the constraints generate an ideal in the tensor
product algebra $g\ell^{x} (n+1) \otimes g \ell (n+1) $ (in the
general sense that multiplication by polynomials in $x_i$ is
allowed). In terms of the theory of the constrained systems it means
that constraints $\varphi_i$ are in involution, i.e. can be
consistently set equal to zero. We can construct the factor (coset)
algebra, this means that the constraints can be imposed preserving
the algebra relations. With the constraints we have $ L^{x} (u+ 1)
L^{(2)} (u) = [u] \ L^{(12)} (u+ 1) $ and both Lax expressions obey
the Yang-Baxter relation (\ref{RLL}).

The constraints can be used to eliminate the generators $E_{i, n+1}$.
After this first reduction we establish another ideal
generated by $E_{n+1, j}$ now with respect to the reduced algebra.
Indeed, we have
$$ [E_{n+1, i} , E_{n+1,j} ] = \delta_{i, n+1} E_{n+1, j} - \delta_{n+1,j}
E_{n+1, i} $$ We may restrict to $ i,j = 1, ..., n$, then the right
hand side is just zero.

The commutators of the tensor product generators with these
constraints are
$$ [E^{x}_{ij} + \tilde E_{ij}, E_{n+1, k} ] = - \delta_{ik} \tilde
E_{n+1,j} $$ The notation $\tilde E_{ij} $ means that in the case
$j=n+1$ the substitution according to the constraint $\varphi_i = 0
$ has to be done. Again the constraints $E_{n+1, i}, i= 1, ..., n$
can be imposed without disturbing the original Lie algebra relations
or the Yang-Baxter relations.

In this way we have eliminated in two steps
the generators $E_{i,n+1}, i= 1, ..., n+1 $ and $E_{n+1, j},
j= 1, ..., n$.
The second tensor factor is reduced to the algebra $g\ell (n)$.

It is useful to draw the attention to the following
point concerning the second reduction step. If one would
consider the constraints $E_{n+1, i} = 0, i=1, ..., n+1$
before the first reduction one would find an obstacle.

No problem arises in the Borel subalgebra involving the
generators $E_{i,j}, i \ge j $. However, the commutators
with the other generators result in terms not vanishing with
these constraints, i.e. preventing from imposing the constraints.

$$ [E_{n+1,i}, E_{j,k} ] = \delta_{ij} E_{n+1,k}
- \delta_{k,n+1} E_{ji} $$

The unwanted second term does not appear if $E_{j, n+1}$
is replaced by a linear combination of $ E_{j s} $,

$$ [E_{n+1,i}, \tilde E_{j,n+1} ] = \delta_{ij} \tilde E_{n+1,k},
\ \ \ \tilde E_{j,n+1} =  \sum_1^n A_s E_{j s} $$

At this point the coefficients are arbitrary. Their relation to
the operators in the other tensor factor arises because this
replacement of $E_{j, n+1}$ should be constructed by
consistent reduction.

\subsection{q-deformed case}

Let us look first how the last  discussed point about the
second reduction appears in the deformed case.
The constraints $E_{n+1, i} = 0, i=1, ..., n+1$
can be consistently imposed within the Borel subalgebra where
these operators belong to. However, the commutation relations
\be
\label{En+1j}
 [E_{n+1,j},  E_{j,n+1} ] =  [E_{n+1} - E_j]
\ee
are not compatible with these constraints.

We write a set of commutation relations of the deformed algebra
in terms of the Lax matrix elements
\be
\label{qEq}
 L_{ji} (0) = {\cal E}_{ij} =
q^{\pm \half E_j} E_{ij}  q^{\pm \half E_i}
\ee
where the sign $+$ stands if $i>j$ and the sign $-$ if
$i<j$.
\be
\label{EE}
 [ {\cal E}_{ij}, {\cal E}_{jk} ] = q^{\pm E_j} {\cal E}_{ik}
\ee
Here the sign $+$ stands if $ i>j>k$ or $i<j, j>k $ and
the sign $-$ stands if $i<j<k$ or $i>j, j<k $.
In Appendix C we outline proofs of (\ref{EE}).

The above relation (\ref{En+1j}) can be rewritten
 replacing  l.h.s. by
$[{\cal E}_{n+1,j}, {\cal  E}_{j,n+1} ] $.
Now, similar to the undeformed case, one observes that the
problem is removed, if ${\cal  E}_{j,n+1} , j =1,..,n+1 $
are replaced by linear combinations as
\be
\label{rep}
 \tilde {\cal  E}_{j,n+1} = \sum_1^n A_s {\cal  E}_{j,s}.
\ee
Indeed,  all terms appearing if calculating the commutator
with the above relations (\ref{EE}) for ${\cal E}_{ij}$
are proportional to some ${\cal E}_{n+1,k} $.
Therefore, the obstacle for imposing the constraints
$ E_{n+1,i}=0 $
is removed if this replacement can be done consistently.

On the other hand we shall see now that this replacement
corresponds to the constraints emerging in the first step
with particular coefficient $A_s$ related to the first
tensor factor.

Consider the relation (\ref{rel}). The second term on r.h.s.
has the form
\be
\label{rest}
 L_r = L^x (1) L (0) = D^{(x) -1} D^{N} M_1 D^{(x)} \ L(0)
\ee

The diagonal matrices $D^x, D^N $ are defined above
(\ref{Dx}, \ref{degen})
and $M_1$ is the $ n+1 \times n+1 $ matrix with all elements
equal  to 1. This term $L_r$ can be written in terms of
\be
\label{phiq}
\varphi_i = \sum_1^{i-1}  X_s {\cal E}_{i s} + X_i \ [E_i]
+ \sum_{i+1}^{n+1}  X_s {\cal E}_{i s}
= \sum_1^{n} X_s {\cal E}_{i s}
\ee
Here we have introduced  $X_i =   q^{-N_{i, n+1}} x_i  $ obeying
$ X_i X_j = q X_j X_i $ for $ n \ge i > j$ and
$X_{n+1 } = x_{n+1}$. ${\cal E}_{ij}$ are defined
in (\ref{qEq}). The second form in (\ref{phiq}) is a short-hand
notation assuming ${\cal E}_{ii} = [E_i]$.

The matrix elements of (\ref{rest}) are
\be
\label{restij}
 (L_{ r})_{ij} = \left (L^x (1) L (0) \right )_{ij} =
q^{N_{i,n+1}}  x_i^{-1} [N_i] \varphi_j
\ee

The algebraic relations involving $\varphi_i, i=1, ..., n+1$
can be derived  immediately  from the fact that both the l.h.s.
and the first term in the r.h.s of (\ref{rel}) obey
the Yang Baxter RLL relation. Therefore,
$$ \check R_{12} (u-v) \left ( [u] L^{12}_1(u+1) L_{r 2} +
[v] L_{r 1} L_2^{12} (v+1) +  L_{r 1} L_{r 2} \right ) =
$$ $$
\left ( [v] L^{12}_1(v+1) L_{r 2} +
[u] L_{r 1} L_2^{12} (u+1) +  L_{r 1} L_{r 2} \right )
\check R_{12} (u-v).
$$
All relations contained here consist only of terms
linear or bilinear in $\varphi_i $, and no term that would not
vanish with $\varphi_i$ is involved.
Therefore the constraints $\varphi_i =0, i=1, ..., n+1 $ can be
imposed preserving the original algebra relations.
The analogous consequence of the RLL relation applies also 
in the undeformed case and leads to the relations 
(\ref{ideal})
derived in the previous subsection in another way.

These constraints are imposed as the first reduction step
and used to replace $ {\cal  E}_{i,n+1}, i= 1, .., n+1 $.
After this one can impose the constraints
$E_{n+1,i}=0, i = 1,..,n$ as the second reduction step.
In this way the second tensor factor being a representation
 of
$U_q(g\ell(n+1))$ is reduced to $U_q(g\ell(n))$.

This procedure results in the iterative construction of 
 $U_q (g\ell (n+1))$ representations from $U_q (g\ell_q (n))$ 
representations by combining the
latter with the special Jordan-Schwinger representations
of $U_q (g\ell (n+1) )$ which can be formulated
in terms of the Heisenberg pairs $x_i, \dd_i, i=1, ..., n+1$.
We have shown that this construction is conveniently formulated
in terms of the Lax matrices, representing the algebras
in question owing to the Yang-Baxter relation.
In particular the reduced tensor product Lax matrix
$L^{(12^\prime )}$ represents the resulting $U_q (g\ell (n+1))$
representation.

Here the reduction eliminated the Cartan-Weyl generators of the
second tensor factor with indices $(n+1,i)$ or $(i, n+1)$, i.e.
referring to the last row and last column of the Lax matrix. 
Obviously other versions can be obtained by choosing for elimination 
the generator related to the row and column of number $i$.

\section{Factorisation}
\setcounter{equation}{0}

After the reduction the relation (\ref{rel}) takes the form
\be
\label{relr}
 L^{x} (u+ 1) L^{\prime} (u) = [u] \
L^{(12^{\prime})} (u+ 1)
\ee
The Lax matrix of the co-product is modified to
$ L^{(12^{\prime})} $ by
substituting zero for the generators of the second tensor factor
with the first index equal to $n+1$,
$E_{n+1, i} = 0 , i=1, ..., n+1 $ and substituting
the generators with the second index equal to $n+1$
according to the constraints $\varphi_i = 0$ by expressions in terms
of the remaining generators.
The same substitution also
modifies the Lax matrix of the second factor. As the result
$ L^{\prime} (u) $ has zeros on  the last
column besides of the lowest entry,
$(L^{\prime} (u))_{n+1,n+1} = [u] $. The other elements
of the last row
are to be calculated according to
the constraints $\varphi_i =0$ in terms of the generators of
the remaining $U_q (g\ell (n))$,
$$
(L^{\prime} (u))_{n+1,i} = q^{-u} \tilde {\cal E}_{i,n+1} =
 - q^{-u} x_{n+1}^{-1} \sum_1^n X_s {\cal E}_{i,s} $$
The remaining $n \times n $ block in $L^{\prime} (u)$
 coincides with  the Lax matrix of a generic $U_q (g\ell (n) ) $
representation.

In the second section we have obtained factorized representations
of the Jordan-Schwinger Lax matrix $L^{(x)} (u) $ in terms
of triangular matrices. The central factor involves
only $N_i = x_i \ \dd_i $ and $x_i$ enter the left and
right factors.

We shall see that the reduced tensor product Lax operator
allows triangular factorized representations.
This follows from the factorisation properties of
the Jordan-Schwinger representations, the reduced relation
(\ref{relr}) and from
$$
L^{\prime} (0) =  m_x^{-1}  L^{n} (0) m_x, \ \ \ 
m_x = D^{(x) -1} m_1  D^{(x) }.
$$
$m_1 $ denotes the $n+1 \times n+1 $ matrix with diagonal and
last row elements equal to $1$ and the other elements zero.
$L^{n} (0)$ denotes the  $ n+1 \times n+1 $ matrix with
zeros on the last row and the last column and the
remaining $n \times n$ block matrix coinciding with
the generic $U_q (g\ell (n)) $ Lax matrix at $u=0$.

\subsection{Undeformed case}

Because of the simple dependence on the spectral parameter $u$
we have in this case
$$
L^{\prime} (u) = m_x^{-1}  L^{n} (u) m_x,\qquad m_x = D^{(x)-1} m_1
D^{(x) }
$$
$$ L^{\prime} (u) = u I + L^{\prime} (0), \qquad
L^{n} (u) = u I + L^{n} (0) $$

The factorization of the Jordan-Schwinger representation
Lax matrix (\ref{LaxJS}) simplifies to
$$
L^{x} (u+1) = m_x^{-1} K(u+1) m_x
$$
$$ K (u+1) = D^{(x) -1} \tilde K(u+1) D^{(x) }
 $$
The Lax matrix of the reduced tensor product is proportional
to the product of $ L^{x} (u+1) $ and $ L^{\prime} (u) $
and therefore factorises as well,
$$
u L^{(12^{\prime})} (u+ 1) =
m_x^{-1} K (u+1) \ L^{n} (u) m_x
$$
The resulting factorisation formula provides a compact formulation
of the algebraic indiction. The algebra $g\ell (n)$
is represented by $L^n (u)$, because its upper block
is the corresponding Lax matrix. The other factors are the
ones contained in the triangular factorisation of the
Jordan-Schwinger form $g\ell (n+1)$ Lax matrix.
$L^{(12^\prime)} (u)$ represents the constructed
algebra $g\ell (n+1)$.

\subsection{q-deformed case}

The non-trivial dependence on the spectral parameter
can be represented as
$$ \lambda L(u) = q^u L_+ - q^{-u} L_- .$$
Applied to the reduced Lax matrix $L^{\prime} (u) $,
$L_+ $ reduces to $L_+^{n} $, the $L_+$ of the $U_q (g \ell (n))$
case supplemented with the n+1st row and n+1st column of zeros.
Let $L_-^{n}$ be the corresponding $L_-$ of  
$U_q (g \ell (n))$
supplemented by zeros in the same way.
However, instead of the latter we need to substitute
$$
 L_-^{\prime} = m_x^{-1}
 \left ( L_-^{n} - L_+^{n} \right )  m_x   + L_+^{n},
\ \ \
m_x = D^{(x)-1} m_1 D^{(x)}
$$
and obtain
\be
\label{Lprime}
\lambda L^{\prime}(u) = (q^u - q^{-u} )  L^{n}_+
+  m_x^{-1}
 q^{-u}\left ( L_+^{n} - L_-^{n} \right )   m_x
\ee
Here the diagonal matrices in the definition of $m_x$ are the
ones introduced for $q \not = 1$ in (\ref{Dx}).

The factorization of the Jordan-Schwinger representation
Lax matrix  now  involves additionally the diagonal matrices
$D_L$ and $D_R$ (\ref{DLR}).

$$
L^{x} (u+1) = D_L^{-1}(u+1)  m_x^{-1}  D_L (u+1)  K (u+1)
D_R^{-1} (u+1)  m_x   D_R (u+1),
$$
$$ K (u+1) = D^{(x) -1} \tilde K(u+1) D^{(x)}.
 $$
The matrices $D_{L/R}$ (\ref{DLR})  depend on the spectral
parameter $u$. In particular, $D_R (u+1)$ has $q^{-2u}$ on the
diagonal besides of the last diagonal element, which is $1$.
In order to get the left factors in the second term for
$L^{\prime}(u)$ closer to the right factors in
$L^{x} (u+1)$ we transform (\ref{Lprime}) to
$$
\lambda L^{\prime}(u) = (q^u - q^{-u} )  L_-^{n}
+ D_R^{-1} (u+1) m_x^{-1}  D_R (u+1)
 q^{u} \left( L_+^{n}-L_-^{n} \right )
D_R^{-1} (u+1)  m_x  D_R (u+1)
$$
$$
=(q^u - q^{-u} )  \left ( L_-^{n} -
D_R^{-1} (u+1)  m_x^{-1}  D_R (u+1) L_-
D_R^{-1} (u+1) m_x  D_R (u+1) \right )
$$ $$
+ D_R^{-1} (u+1)  m_x^{-1}  D_R (u+1)
 L^{n \prime} (u)
D_R^{-1} (u+1)  m_x D_R (u+1)
$$
$ L^{n \prime} (u) $ is the Lax matrix of $U_q(g\ell (n))$
supplemented by zeros in the n+1st row and n+1st column.
Notice that the first term in the last expression
has non-vanishing elements only on the last row.
This allows to rewrite the sum into
one factorised expression in terms of
$L^{n} (u) $ which differs from $ L^{n \prime} (u) $
in the last diagonal element $(n+1,n+1)$ being now non-zero
and equal to $[u]$.
\be
\label{Lprimeq}
L^{\prime}(u) =
M_R^{-1} (u+1)  L^{n} (u) M^{\prime} M_R (u+1),
\ee
$$
M_R (u+1) = D_R^{-1} (u+1) m_x D_R (u+1), \ \ 
m_x = D^{(x)-1} m_1 D^{(x)}
$$
$$
M^{\prime} = I + \left ( L_-^{n} -
 M_R^{-1}(u+1)  L_-
M_R (u+1) \right ),
$$
With this factorised form of $L^{\prime}(u)$ it is now
straightforward to write the factorization of the reduced tensor
product Lax  matrix.

\begin{prop}
The Lax matrix  $L^{(12^{\prime})} (u+ 1)$ of a 
 $U_q (g\ell (n+1))$ representation can be constructed from the
Lax matrix of a  $U_q (g\ell (n) )$ representation
and $n+1$ Heisenberg
conjugated pairs $x_1, \dd_i, i=1, ..., n+1$ as
\be
\label{Lfactq}
[u] L^{(12^{\prime})} (u+ 1) =
M_L (u+1) K(u+1) L^{n} (u) M^{\prime} M_R (u+1)
\ee
$L^n (u)$ is block diagonal with the upper $n \times n$ block
being the Lax matrix of $g\ell_q (n)$ and the last diagonal
element equal to $[u]$.

$M_L, M_R$ are lower-triangular, $K(u)$
is upper-triangular and they appear as factors
in the Lax matrix of the Jordan-Schwinger form
of $U_q (g\ell (n+1) )$ constructed in terms of $x_i \dd_i$,
\bea
L^x(u)=  M_L(u) K(u) M_R (u) \cr
M_L(u) = D_L^{-1} (u) m_x^{-1} D_L (u),  \ \
M_R(u) = D_R^{-1} (u) m_x D_R (u), \cr
m_x = D^{(x)-1}(u) m_1 D^{(x)} (u), \ \
K(u) = D^{(x) -1}(u) \tilde K(u) D^{(x)} (u)
\eea
$\tilde K$ is given in (\ref{tildeK}) and the diagonal matrices
$D^{(x)}, D_L, D_R$ are defined in (\ref{Dx}, \ref{DLR}).
$m_1$ is lower triangular with elements equal to 1
on the diagonal
and on the lowest row and all other elements zero.
$M^{\prime} $ is lower-triangular with units on the diagonal
and the only other non-vanishing elements on the last row.
It is calculated from the lower-triangular part $L_-$
of the $U_q (g\ell (n))$ Lax matrix and $D_L, D_R, D^{(x)}$
as in (\ref{Lprimeq})
\end{prop}

An alternative factorised form can be obtained where
the analoga of $K$ is lower triangular and of
$M_L, M_R, M^{\prime}$  are upper triangular.
Also the block structure of the analogon of $L^{n} (u)$
is opposite with the $U_q (g\ell (n))$ Lax matrix appearing
as the lower $n \times n$ block.
One arrives at this alternative form if one uses the constraints
$\varphi_i =0$ to eliminate $E_{i,1}$ and proceeds with the
constraints $E_{1,i} =0 $ to do the reduction to 
$U_q(g\ell (n))$
in a different way. The alternative factorisation
of the Jordan-Schwinger Lax matrix interchanging
the roles of upper and lower triangular matrices
is then applied.
Further forms exist corresponding to the reduction
by consistent elimination of the row and column
of number $i$.

\section{Discussion}

The Lax matrices of the Jordan-Schwinger type representations of
$g\ell(n+1)$ or $U_q(g\ell(n+1))$ show a simple 
structure allowing useful factorised expressions. Although this
form covers only a special class of representations 
it can be used as building block
for constructing generic representations by the
 method of algebraic induction. A generic representation is 
obtained by combining a Jordan-Schwinger type representation 
formulated in terms of $n+1$ Heisenberg pairs
with a generic representation of the corresponding algebra with
rank lower by one unit. 
The representation parameter $\ell$ associated
with the Jordan-Schwinger representation becomes the
additional weight component. Iterating this procedure
a generic representation is finally constructed in terms
of $\half (n+1)n $ Heisenberg pairs; $n$ pairs of them are
eliminated by solving the corresponding representation 
constraints, specifying simultaneously the $n$ weight components
as the representation labels of $s\ell(n+1)$ or of its 
trigonometric deformation.

In this paper we have shown how the algebraic induction 
is derived  from the relation between the product of Lax 
matrices of two representations and the Lax matrix composed
of the co-product generators. We have specified one of these 
representations in the Jordan-Schwinger form and made use of the
simple factorisation properties of the latter. The other
representation in the tensor product can be constrained to
a representation of the corresponding algebra with reduced 
rank while preserving the algebra relations in the
tensor product representation. 
In this way the constructed representation is formulated 
in terms of the
constrained tensor product Lax matrix. The relation of the 
latter to the product of Lax matrices of the two 
representations, one being of the Jordan-Schwinger type
and the other generic but reduced in rank, results in
factorised expressions.

The factorised expression (\ref{Lfactq})
provides a short and simple
formulation of the iterative construction of
$U_q (g\ell (n+1))$ representations from 
$U_q(g\ell (n))$ representations 
equivalent to the involved expressions of the
algebraic induction known so far. We see that the
Lax matrices provide the appropriate formulation.

We have pointed out that there are several forms of 
factorisation of the Jordan-Schwinger Lax matrix and also of the
constructed generic representation Lax matrix. The form
considered explicitely here selects the last row and last 
column of the matrices. Correspondingly the parameter $\ell$
in the representation constraint becomes the $n$th component of the
weight of the constructed representation. 
The other forms, selecting instead 
the row and the column of number $i$, may be of interest 
because the comparison of different forms results in 
explicit representations of the intertwining operators
relating the  equivalent representations 
differing in the ordering of the weight components.  
Explicit intertwining operators are of interest in the
factrisation method of constructing the Yang-Baxter 
$R$ operator for generic representations.

\section*{Acknowledgement}

This work has been supported by  RFFI grant 07-02-92166,
RFFI grant 08-01-00638,
 DFG grant 436 Rus 17/4/07 (S.D.),
by DFG grant 436 Arm 17/1/07, by Volkswagen Stiftung (D.K.),
by RFFI grant 06-01-00186-a (P.V.), 
by DFG grant KI 623/5-1 and NTZ of Leipzig University.
One of us (P.V.) is grateful to the 
" Dynasty " foundation and to DAAD for support.

%%%%%%%%%%%%%%%%%%%%%%%%%%%%%%%%%%%%%%%%%%%%%%%%%%%%%%%%%%%%%%%%%%%%

\section*{Appendix A}
\setcounter{equation}{0}

We define the co-product on the generators in the Chevalley form
as in (\ref{co}).
The Cartan-Weyl generators are defined iteratively as (\ref{iter})
for the generators on both tensor factors 
$E^{(1)}_{ij}, E^{(2)}_{ij}$ and and the ones on the
the tensor product $E^{(12)}_{ij}$.
We intend to write the co-product explicitly in the 
Cartan-Weyl basis.
In the first step we obtain
$$E^{(12)}_{i, i+2} = E_{i,i+2}^{(1)} q^{\half N_{i+2}^{(2)} 
- N_i^{(2)})}
+ $$
$$
E_{i,i+2}^{(2)} q^{- \half (N_{i+2}^{(1)} -N_i^{(1)})}
+ (q^{-\half}-q^{\frac{3}{2})})
q^{ \half (N_{i+2}^{(2)} - N_{i+1}^{(2)} -N_{i+1}^{(1)} + N_i^{(1)} )}
E_{i+1,i+2}^{(1)}   E_{i,i+1}^{(2)}
$$
The generic case is obtained as
 \be \label{coCW-}
\!E^{(12)}_{i, i+k}\! =\! E_{i,i+k}^{(1)} q^{\half N_{i+k}^{(2)}\!
-\! N_i^{(2)})}\!+\! E_{i,i+k}^{(2)} q^{-\half (N_{i+k}^{(1)}
-N_i^{(1)})}\! -\! \lambda \sum_{s=1}^{k-1}
 E_{i+s,i+k}^{(1)} q^{ \half (N_{i+k}^{(2)} - N_{i+s}^{(2)} -N_{i+s}^{(1)}
+ N_i^{(1)} )} E_{i,i+s}^{(2)} \ee The iteration in the direction of
$i>j$ results in
 \be \label{coCW+}
\! E^{(12)}_{j+k,j}\! =\! E_{j+k,j}^{(1)} q^{\half (N_{j+2}^{(2)}
-N_{j}^{(2)}}\!+\! E_{j+k,j}^{(2)} q^{-\half (N_{j+2}^{(1)}
-N_{j}^{(1)}}\! +\! \lambda \sum_{s=1}^{k-1}  E_{j+s,j}^{(1)} q^{
-\half (N_{i+k}^{(2)} - N_{i+s}^{(2)} -N_{i+s}^{(1)} + N_i^{(1)} )}
E_{j+k,j+s}^{(2)} \ee

The Lax matrix (\ref{Lax}) can be decomposed as
$$ \lambda L(u) = q^{u} L_+ - q^{-u} L_- , \lambda = q - q^{-1} $$
where $L_+$ is upper triangular and $L_-$ is lower triangular with the
elements in terms of the Cartan-Weyl generators,
$$ (L_-)_{i,j} = -\lambda q^{-\half N_i} \ E_{j,i} \ q^{-\half N_j} ,\qquad i > j, $$
$$ (L_-)_{i.i} = q^{-N_i} $$
$$ (L_+)_{i,j} = \lambda q^{\half N_i} \ E_{j,i} \ q^{\half N_j} ,\qquad i < j, $$
$$ (L_+)_{i.i} = q^{N_i} $$

We rewrite the result for $E^{(12)}_{i,i+k}$ in order to obtain the
corresponding relation for the Lax matrix elements.
$$ q^{-\half(E^{(1)}_{i+k} + E^{(2)}_{i+k})} E^{(12)}_{i,i+k}
q^{-\half(E^{(1)}_{i} + E^{(2)}_{i})} =
\left ( q^{-\half E_{i+k}^{(1)} } E^{(1)}_{i,i+k} q^{-\half E_{i}^{(1)} }
\right ) q^{-E_i^{(2)} } + $$ $$ q^{-E_{i+k}^{(1)} }
\left ( q^{-\half E_{i+k}^{(2)} } E^{(2)}_{i,i+k} q^{-\half E_{i}^{(2)} }
\right )
- \lambda \sum_{s=1}^{k-1}
 \left ( q^{-\half E_{i+k}^{(1)} } E^{(1)}_{i+s,i+k} q^{-\half E_{i+s}^{(1)} }
\right )
\left ( q^{-\half E_{i+s}^{(2)} } E^{(2)}_{i,i+s} q^{-\half E_{i}^{(2)} }
\right )
$$
Up to a factor $- \lambda^{-1}$ this coincides term by term with
$$ (L^{(12)}_-)_{i+k,i} = \sum_{s=0}^{k} (L_-^{(1)})_{i+k, i+s}
(L_-^{(2)} )_{i+s, i} $$
The case $E^{(12)}_{j+k,j}$ is analogous. In this way we have checked
that the relations (\ref{Lax12})
$$ L_{\pm}^{(12)} = L_{\pm}^{(1)} L_{\pm}^{(2)} $$
are indeed equivalent to the coproduct rules (\ref{co}).

\section*{Appendix B}

First of all the constraint (\ref{repc}) fixes the 
representation of the $g\ell(1)$ subalgebra generated by
$\sum_1^{n+1} N_i$. The constraint  allows to eliminate one pair,
e.g. $x_{n+1}, \dd_{n+1}$. It is  solved for 
$N_{n+1}$ and the representation is restricted to functions of 
${x_i \over x_{n+1} }$, so we can set $x_{n+1} = 1$ for
simplicity. We have a lowest weight module spanned by polynomials 
of $x_i, i=1, ..., n$. The constant 1 is the lowest weight 
vector, in particular an eigenvector of 
$h_i = \half (N_i - N_{i+1})$, and the weight components are
$(0, ...,0,\ell)$. 

In the situation after the constraint has been imposed
the matrix elements with indices $i,j = 1, ...,n$ are
still given by (\ref{LaxN}) and the  remaining ones are
$$ L_{n+1, n+1} = [u+ 2 \ell - \sum_1^n N_s ], \ \qquad
L_{i, n+1} = q^{u- \half + \tilde N_{i, n+1} } {1 \over x_i} [N_i],
$$  $$ L_{n+1, j} = q^{-u + \half - \tilde N_{n+1,j} } x_j [2 \ell-
\sum_1^n  N_s], \ \qquad \tilde N_{i,n+1} = \tilde N_{n+1,i} = \ell
- \half \sum_1^{i-1} N_s + \half \sum_{i+1}^n N_s
$$

We observe the simplification of the Lax matrix by
$$ D_{\ell}^{((x)) } L (u) D_{\ell}^{(x) -1}  =
\tilde L_{\ell} (u) $$
$$D^{(x)}_{\ell} = {\rm diag}
( q^{-\tilde N_{1,n+1}} x_1, ..., q^{-\tilde N_{n,n+1}} x_n,
q^{-\half} )
$$
Here $\tilde L_{\ell}$ coincides with the simplified Lax matrix
$\tilde L$ (\ref{Ltilde}) above besides of the
substitution

\be
\label{tildeNn+1}
 N_{n+1} \rightarrow \tilde N_{n+1} =
2\ell +1 - \sum_1^n N_s.
\ee

The remaining steps are now the same as above, only the latter
substitution has to be done. Thus $\tilde M_R $ is unchanged
and the substitution turns $\tilde M_L$ to $\tilde M_L^{(\ell)}$ and
$\tilde K$ to $\tilde K_{\ell}$. The upper traingular matrix
$\tilde K_{\ell} $ coincides (up to the substitution)
(\ref{tildeNn+1}) with $\tilde K$ (\ref{tildeK})
up to the lowest diagonal element, which turns to
$$ (\tilde K_{\ell} )_{n+1,n+1} = [u+ 2\ell] q^{-\sum_1^n N_s }
$$

Notice that the representation parameter dependence in
$D_{\ell}^{(x)} $ can be easily absorbed by rescaling
$ q^{-\ell} x_i \rightarrow x_i,   i=1, ...,n$,
$$ D_{\ell}^{(x)} \rightarrow D_1^{(x)} =
{\rm diag} (..., q^{\ell -\tilde N_{i,n+1}} x_i, ..., 1)
\cdot q^{-\half }
$$
leaving $N_i$ unchanged. Further we write
$$ \tilde K_{\ell}  = D_{1 \ell} \tilde K_1, \ \ \
D_{1 \ell} = {\rm diag} (1, ..., 1, [u+2\ell ]  ) $$
Then all the remaining representation parameter dependence
resides in $D_{1\ell}$ and $D_L^{(\ell)}$ and
 is factorized to the left:
$$ L (u) = M_L (u) D_{1\ell} K_1(u) M_R (u) $$
$$ K = D_1^{(x)-1} \tilde K_1 D_1^{(x)},\qquad
M_{L/R} = D_1^{(x)-1} \tilde M_{L/R} D_1^{(x)} $$
$$ \tilde M_R = D_R^{-1} (u) m_1 D_R (u),\qquad
\tilde M_L = D_L^{(\ell) -1} (u) m^{-1}_1 D^{(\ell)}_L (u)
$$
Therefore the results for reduction and factorization are not
changed essentially besides of the same modification of the
left-most factors.

\section*{Appendix C}

The algebra relations of $U_q (g\ell (n+1)) $ 
are implicit in the
Yang-Baxter RLL relation (\ref{RLL}). Explicitly we have for the
Lax matrix
$$ \lambda L(u) = q^u L_+ - q^{-u} L_- $$
\bea
\left ( L_- \right )_{ij} =
\left \{
\begin{array}{cc}
0, & i<j \\
q^{-E_i}, &     i=j,  \\
- \lambda {\cal E}_{ji}, &     i>j \\
\end{array}
\right. {\ } \ \ \ \ \ \left ( L_+ \right )_{ij} = \left \{
\begin{array}{cc}
\lambda {\cal E}_{ji}, & i<j \\
q^{E_i}, &     i=j \\
0, &     i>j \\
\end{array}
\right. {\ } \eea where $ {\cal E}_{ij} = q^{\mp E_j} E_{ij} q^{\mp
E_i} $ for $i<j$ or $i>j$, respectively. Further, for the
fundamental R matrix we have
$$\lambda \check R_{12} (u) = q^u \check R_+ - q^{-u} \check R_-,
\check R_-|_q  = \check R_+ |_{q^{-1}} = (\check R_+|_q)^{-1} $$
$$ \left ( \check R_+ \right )^{i_1 i_2}_{j_1 j_2} =
\delta^{i_1}_{j_2} \delta^{i_2}_{j_1} q^{\delta_{i_1 i_2} } +
\lambda  \delta^{i_1}_{j_1} \delta^{i_2}_{j_2} \theta (i_1 - i_2) $$
By separating the dependence on $u+v$ and $u-v$ the spectral parameter
dependent $RLL$ relation implies
$$ \check R_+ L_{1 \pm} L_{2 \pm} = L_{1 \pm} L_{2 \pm} \check R_+ ,$$
$$ \check R_+ L_{1 +} L_{2 -} = L_{1 -} L_{2 +} \check R_+ .$$
As an example we pick up the case with all subscripts $+$,
$$ (\check R_+)^{i_1 i_2}_{k_1 k_2}  L^{k_1}_{ + j_1}
L^{k_2}_{+ j_2 } = L^{i_1}_{+ k_1} L^{i_2}_{+ k_2 } (\check R_+)^{k_1
k_2}_{j_1 j_2} $$
and substitute the explicit form of $R_+$. In the special case
$i_2 < i_1=j_1 < j_2$ we find
we find
$$ \lambda^2 [ {\cal E}_{i_1 i_2}, {\cal E}_{j_2 i_1}] +
\lambda q^{E_{i_1}} \lambda {\cal E}_{j_2 i_2} = 0
$$

By relabeling indices this results in the corresponding relation of
(\ref{EE}).

The Jordan-Schwinger representation provides an alternative of checking
algebra relations. Starting from (\ref{EJ}) we derive easily
$$ [E_{ij}, E_{jk} ]_q = E_{ik},\qquad i<j<k, $$
$$ [E_{ij}, E_{jk} ]_{q^{-1}} = E_{ik},\qquad i>j>k, $$
$$ [E_{ij}, E_{ji} ]_{1} = [N_i - N_j],\qquad i \not = j, $$
$$ [E_{i+1,i}, E_{i,j} ]_1 = E_{i+1,j}
q^ {N_i - N_{i+1}},\qquad i+1 <j, $$
$$ [E_{i-1,i}, E_{i,j} ]_1 = E_{i-1,j}
q^ {N_{i-1} - N_{i}},\qquad i-1>j,
$$
 \be \label{CWrel} [E_{k,i},
E_{i,i+1} ]_1 = E_{k,i+1} q^ {N_{i+1} - N_{i}},\qquad k-1>i.
 \ee

The proofs in the Jordan-Schwinger form rely on
the relations (\ref{EJ}). Let us do the 4th relation
as an example.
$$ [E^{x}_{i+1,i}, E^{x}_{i,j} ]_1 =
[E^{J}_{i+1,i}, q^{-\sum_{i+1}^{j-1} N_s} E^{J}_{i,j} ]_1
 = [E^{J}_{i+1,i}, E^{J}_{i,j} ]_{q^{-1}}
q^{-\sum_{i+1}^{j-1} N_s}$$ $$
= q^{N_i} E^{J}_{i+1,j} q^{-\sum_{i+1}^{j-1} N_s} =
E^{x}_{i+1,j} q^{N_i - N_{i+1} }
$$
Relying on the latter relations we can check (\ref{EE}) For example,
let $ i<j<k$
$$[{\cal E}_{ij}, {\cal E}_{jk} ]_1 =
q^{-E_j} q^{-\half E_k} [ E_{ij}, E_{jk} ]_q q^{-\half E_i} =
- q^{-E_j} {\cal E}_{ik}. $$
In the last step the 1st relation of (\ref{CWrel}) has been applied.


\vspace*{1cm}
\begin{thebibliography}{99}



\bibitem{GN} I.M. Gelfand and M.A. Naimark,
Unitary representations of the classical groups,

Trudy Math. Inst. Steklov, Vol. 36, Moscow-Leningrad 1950.
(German translation: Akademie Verlag, Berlin 1957)

\bibitem{BW}
A. Borel and A. Weil, Representations lineaires et espaces homogenes
K\"ahlerians des groupes de Lie compactes, Sem. Bourbaki, May 1954,
(expose J.-P. Serre).

\bibitem{H}
K.T. Hecht, The vector coherent state method and its application
to problems of higher symmetry, Lecture Notes in Physics 290,
Springer, 1987.

\bibitem{HP}
T. Holstein and H. Primakoff,
``Field dependence of the intrinsic domain magnetization of a ferromagnet,
``
Phys. Rev. {\bf 58} (1940), 1098 - 1113.

\bibitem{BL}
L.C. Biedenharn and M.A. Lohe,
``An extension of the Borel-Weil construction to the
quantum group $U_q(n)$ ``, 
Commun. Math. Phys. {\bf 146} (1992) 483 - 504;

Quantum group symmetry and q-tensor algebras,
World Scientific 1995.



%\cite{Dobrev:1994ne}
\bibitem{Dobrev:1994ne}
  V.~K.~Dobrev, P.~Truini and L.~C.~Biedenharn,
  ``Representation theory approach to the polynomial solutions of q
  differenceequations:U-q(sl(3)) and beyond,''
  J.\ Math.\ Phys.\  {\bf 35} (1994) 6058
  [arXiv:q-alg/9502001];
  %%CITATION = JMAPA,35,6058;%%
%\cite{Dobrev:1996rw}
%\bibitem{Dobrev:1996rw}

  V.~K.~Dobrev and P.~Truini,
  ``Irregular U(q)(sl(3)) representations at roots of unity via
  Gelfand-(Weyl)-Zetlin basis,''
  J.\ Math.\ Phys.\  {\bf 38} (1997) 2631.
  %%CITATION = JMAPA,38,2631;%%


%\cite{Derkachov:2005hw}
\bibitem{Derkachov:2005hw}
  S.~E.~Derkachov,
  ``Factorization of the R-matrix.I,''
  arXiv:math/0503396.
  %%CITATION = MATH/0503396;%%




%\cite{Derkachov:2006fw}
\bibitem{Derkachov:2006fw}
  S.~E.~Derkachov and A.~N.~Manashov,
  ``R-Matrix and Baxter Q-Operators for the Noncompact SL(N,C) Invariant
  Spin Chain,''
  SIGMA {\bf 2} (2006) 084
  [arXiv:nlin/0612003].
  %%CITATION = 00480,2,084;%%




%\cite{Derkachov:2007gr}
\bibitem{Derkachov:2007gr}
  S.~Derkachov, D.~Karakhanyan and R.~Kirschner,
  ``Yang-Baxter R operators and parameter permutations,''
  Nucl.\ Phys.\  B {\bf 785} (2007) 263
  [arXiv:hep-th/0703076].
  %%CITATION = NUPHA,B785,263;%%




%\cite{Awata:1993hj}
\bibitem{Awata:1993hj}
  H.~Awata, M.~Noumi and S.~Odake,
  ``Heisenberg realization for U-q(sl(n)) on the flag manifold,''
  Lett.\ Math.\ Phys.\  {\bf 30} (1993) 35
  [arXiv:hep-th/9306010].
  %%CITATION = LMPHD,30,35;%%


%\cite{Dobrev:1993sr}
\bibitem{Dobrev:1993sr}
  V.~K.~Dobrev,
  ``Q difference intertwining operators for U-q(sl(n)): General setting and
  the case n=3,''
  J.\ Phys.\ A  {\bf 27} (1994) 4841
  [Erratum-ibid.\  A {\bf 27} (1994) 6633]
  [arXiv:hep-th/9405150].
  %%CITATION = JPAGB,A27,4841;%%



%\cite{Jimbo:1985zk}
\bibitem{Jimbo:1985zk}
  M.~Jimbo,
  ``A q difference analog of U(g) and the Yang-Baxter equation,''
  Lett.\ Math.\ Phys.\  {\bf 10} (1985) 63;
  %%CITATION = LMPHD,10,63;%%
%\cite{Jimbo:1985vd}
%\bibitem{Jimbo:1985vd}
%  M.~Jimbo,
  ``A Q Analog Of U (Gl (N+1)), Hecke Algebra And The Yang-Baxter
  Equation,''
  Lett.\ Math.\ Phys.\  {\bf 11} (1986) 247-252.
  %%CITATION = LMPHD,11,247;%%


%\cite{Kulish:1981gi}
\bibitem{Kulish:1981gi}
  P.~P.~Kulish, N.~Y.~Reshetikhin and E.~K.~Sklyanin,
  ``Yang-Baxter Equation And Representation Theory. 1,''
  Lett.\ Math.\ Phys.\  {\bf 5} (1981) 393; 
  %%CITATION = LMPHD,5,393;%%

%\cite{Kulish:1981bi}
%\bibitem{Kulish:1981bi}
  P.~P.~Kulish and E.~K.~Sklyanin,
  ``Quantum Spectral Transform Method. Recent Developments,''
  Lect.\ Notes Phys.\  {\bf 151} (1982) 61.
  %%CITATION = LNPHA,151,61;%%

%\cite{Faddeev:1987ih}
\bibitem{Faddeev:1987ih}
  L.~D.~Faddeev, N.~Y.~Reshetikhin and L.~A.~Takhtajan,
  ``Quantization of Lie Groups and Lie Algebras,''
  Leningrad Math.\ J.\  {\bf 1} (1990) 193
  [Alg.\ Anal.\  {\bf 1} (1989) 178].
  %%CITATION = 00040,1,178;%%

%\cite{Faddeev:1996iy}
\bibitem{Faddeev:1996iy}
  L.~D.~Faddeev,
  ``How Algebraic Bethe Ansatz works for integrable model,''
  arXiv:hep-th/9605187.
Published in *Les Houches 1995, Relativistic gravitation
and gravitational radiation* pp. 149-219
  %%CITATION = HEP-TH/9605187;%%



\bibitem{KS}
A. Klimyk and K. Schm\"udgen, Quantum groups and their representations,
Springer, 1997.

%\cite{Isaev:1995cs}
\bibitem{Isaev:1995cs}
  A.~P.~Isaev,
  ``Quantum Groups And Yang-Baxter Equations,''
  Sov.\ J.\ Part.\ Nucl.\  {\bf 26} (1995) 501;
see also the extended version: 
preprint Bonn (2004) MPI 2004-132.  
  %%CITATION = SJPNA,26,501;%%





\end{thebibliography}
\end{document}